\documentclass[prd]{revtex4}

\usepackage{amsmath}
\usepackage{bm}
\usepackage{graphics}
\usepackage{color}

\def\gsim{\;\rlap{\lower 2.5pt
 \hbox{$\sim$}}\raise 1.5pt\hbox{$>$}\;}
\def\lsim{\;\rlap{\lower 2.5pt

   \hbox{$\sim$}}\raise 1.5pt\hbox{$<$}\;}

\def\be{\begin{equation}}
\def\ee{\end{equation}}
\def\ba{\begin{eqnarray}}
\def\ea{\end{eqnarray}}

\def\trh{T_{\rm{RH}}}
\def\k0{k_0^{\rm{p}}}

\begin{document}

\title{Relic gravitational waves in the frame of  slow-roll inflation with a power-law potential and the detection}

\author{Minglei Tong}

\thanks{Email: mltong@ntsc.ac.cn}

\affiliation{ National Time Service Center, Chinese Academy of Sciences, Xi'an, Shaanxi 710600,  China }
\affiliation{Key Laboratory of Time and Frequency Primary Standards, Chinese Academy of Sciences, Xi'an, Shaanxi 710600,  China }

\begin{abstract}
 We obtained the analytic solutions of relic gravitational waves (RGWs) for the slow-roll inflation with a power-law form potential of the scalar field, $V=\lambda\phi^n$.  Based on a reasonable range of $n$ constrained by cosmic microwave background (CMB) observations, we give tight constraints of the tensor-to-scalar ratio $r$ and the inflation expansion index $\beta$ for the fixed scalar spectral index $n_s$. Even though, the spectrum of RGWs in low frequencies is hardly depends on any parameters, the high frequency parts will be affected by several parameters, such as $n_s$, the reheating temperature $\trh$ and the index $\beta_s$ describing the expansion from the end of inflation to the reheating process. We  analyzed  in detail  all the factors which would affect the spectrum of RGWs in high frequencies including the  quantum normalization.  We found that   the future GW detectors   SKA,   eLISA, BBO and  DECIGO    are promising to catch the signals of  RGWs.
 Furthermore, BBO and DECIGO
 have the potential not only to distinguish the spectra with different parameters but also to examine the validity of the quantum normalization.

\



\end{abstract}

\maketitle

\large

\section{ Introduction}

The validity of general relativity and quantum mechanics make sure the generation of  a stochastic background of relic gravitational waves (RGWs)
\cite{grishchuk1,grishchuk,grishchuk3,starobinsky,Maggiore,Giovannini} during the
early inflationary stage, whose  primordial  amplitude could be determined by
the quantum normalization at the time of the wave modes crossing the horizon during the inflation.
 Since the interaction of RGWs with other cosmic components
is  very weak,  the evolution of RGWs are mainly determined
by the behavior of  cosmic expansion
including the current acceleration \cite{zhang2,Zhang4}.
Therefore, RGWs could serve as an unique tool to study the very early Universe
earlier than the recombination stage when the cosmic microwave background (CMB) radiation was generated. As an interesting source for gravitational wave (GW) detectors,
RGWs exist everywhere and anytime  unlike GWs radiated by usual astrophysical process.
Moreover,
RGWs spread a very broad range of frequency, $10^{-18}-10^{10}$ Hz, making themselves become
  one of the major scientific goals of various GW detectors with different response frequency bands.
  The current and planed GW detectors contain  the ground-based interferometers,
such as  LIGO \cite{ ligo1}, Advanced LIGO \cite{ligo2,advligo},
VIRGO \cite{virgo,virgocurve}, GEO \cite{geo},  KAGRA \cite{KAGRA} and ET \cite{Punturo,Hild} aiming at the frequency range $10^2-10^3$ Hz;
the space interferometers, such as the future eLISA/NGO \cite{lisa0,lisa} which is sensitive in the frequency range $10^{-4}-10^{-1}$ Hz,  BBO \cite{Crowder,Cutler}  and DECIGO
 \cite{Kawamura,Kudoh} which  both are sensitive in the
frequency range $0.1-10$ Hz; and the pulsar timing array, such as  PPTA \cite{PPTA,Jenet} and the planned
SKA \cite{Kramer} with the frequency window $10^{-9}-10^{-6}$ Hz.
Besides, there some potential very-high-frequency GW detectors, such as
the waveguide detector \cite{cruise},
the proposed gaussian maser beam detector around GHz \cite{fangyu}, and the 100 MHz detector with a pair of
75-cm baseline synchronous recycling  interferometers \cite{Akutsu}.
Furthermore,   the very low frequency portion of
RGWs also contribute to the  anisotropies
and polarizations of CMB \cite{basko},
yielding a magnetic type polarization of CMB
as a distinguished signal of RGWs.
WMAP \cite{Peiris,Spergel,Hinshaw,Komatsu},
Planck \cite{Planck}, the ground-based  ACTPol  \cite{Niemack}  and the proposed CMBpol \cite{CMBpol}
are of this type.

The reheating temperature, $T_{\rm{RH}}$, carries rich information of the early Universe, and   relates
to the decay rate of the inflation as $T_{\rm{RH}}\propto \sqrt{\Gamma}$ \cite{Kolb,Nakayama}.
Recently, the temperature of the reheating, $T_{\rm{RH}}$, was evaluated  \cite{Mielczarek} according to  the CMB observations by WMAP 7 \cite{Komatsu}, combining with  the slow-roll inflation scenarios. Furthermore, the resultant RGWs was studied in \cite{Tong6}. However, these pieces of work underwent the assumption of a fixed form of the potential of the scalar filed driven the slow-roll inflation, $V(\phi)=\frac{1}{2}m^2\phi^2$. In this paper, we study a more general case of  $V(\phi)=\lambda\phi^n$ \cite{Turner}, where $\lambda$ is constant.  Moreover, we adopt the limitation $n<2.1$ obtained from the spectrum of CMB \cite{martin}.
For a non-fixed value of $n$, it is hard to evaluate the temperature of the reheating process, $\trh$, using the method employed in \cite{Mielczarek}. Thus, we choose several values of $\trh$ lying in the range of $\sim 10^4-10^8$ GeV, where the lower limit and the upper limit of $\trh$ are obtained from the constraints in \cite{martin} and \cite{Bailly}, respectively. In the text, one will see that $n$ and $\trh$ affect the increases of the scale factor in the early stages of the Universe.
 Once  all the
expansion histories  of different stages are determined,
the evolutions of the RGWs at various phases can   be determined subsequently. For the
present time, the solutions of RGWs can be obtained, whose  different frequency bands correspond  to
the k-modes re-entered the horizon at different phases.
On the other hand, the anisotropies due to the tensor metric perturbations (gravitational waves) can be scaled
to those due to the observations of the scalar perturbations by introducing a parameter $r$ called tensor-to-scalar
ratio. Under the frame of the slow-roll inflation scenario, $r$ will be constrained in a  narrow range due to the constraints from $n$ for a given value  of the scalar spectral index $n_s$. Similarly, the inflation expansion index $\beta$ will also be constrained in a narrow range.
Besides, there is a simple relation between $n$ and the preheating expansion index $\beta_s$ describing the
expansion behavior of the universe from the end of inflation
 to the reheating process. As will be shown below, the RGWs in the very high frequencies are sensitively dependent on the parameters $n_s$, $\beta_s$ and $\trh$. Furthermore, the spectra of RGWs also depends on the condition of the quantum normalization.
To this end, the spectra of RGWs given by  different parameters and different conditions will
confront  the various current and planed GW detectors. The future detectors BBO and DECIGO are quite promising not only  to determine various parameters but also to  examine the validity of the quantum normalization.

   Throughout this paper,  we use the units   $c=\hbar=k_B=1$.
Indices $\lambda$, $\mu$, $\nu$,... run from 0 to 3, and $i$, $j$, $k$,... run from 1 to 3.

\section{   RGWs in the accelerating universe}

In a spatially flat universe, the existence of perturbations modifies the Friedmann-Robertson-Walker metric  to be
\be
ds^2=a^2(\tau)[-d\tau^2+(\delta_{ij}+h_{ij})dx^idx^j],
\ee
where $a(\tau)$ is the scale factor, $\tau$ is the conformal time, and $h_{ij}$ stands for the  perturbations to the  homogenous and isotropic spacetime background. In general, there are three kinds of perturbations: scalar perturbation, vectorial perturbation and tensorial perturbation.
In this paper we only consider the tensorial perturbation, that is, gravitational waves. In the transverse-traceless (TT) gauge,
$h_{ij}$ satisfies:   $\frac{\partial h_{ij}}{\partial x^j}=0$  and $h^i_{\,\,i}=0$, where we used the Einstein  summation convention.
In the Fourier $k$-modes space, it can be written as
\be
\label{planwave}
h_{ij}(\tau,{\bf x})=
   \sum_{\sigma}\int\frac{d^3k}{(2\pi)^{3/2}}
         \epsilon^{(\sigma)}_{ij}h_k^{(\sigma)}(\tau)e^{i\bf{k}\cdot{x}},
\ee
where $\sigma=+,\times$ stands for the two polarization states,
 the comoving wave number $k$ is related with
the wave vector $\mathbf{k}$ by $k=(\delta_{ij}k^ik^j)^{1/2}$,
$h_{-k}^{(\sigma)*}(\tau)=h_k^{(\sigma)}(\tau)$
ensuring  $h_{ij}$ be real, and the polarization
tensor  $\epsilon^{(\sigma)}_{ij}$ satisfies \cite{grishchuk}:
\be
\epsilon^{(\sigma)}_{ij}\epsilon^{(\sigma'){ij}}=2\delta_{\sigma\sigma'}, \quad
\epsilon^{(\sigma)}_{ij}\delta^{ij}=0, \quad
\epsilon^{(\sigma)}_{ij}n^j=0,\quad
\epsilon^{(\sigma)}_{ij}(-\mathbf{k})=\epsilon^{(\sigma)}_{ij}(\mathbf{k}).
\ee
In terms of the mode $h^{(\sigma)}_{k}$,
the wave equation is
\be \label{eq}
h^{ (\sigma) }_{k}{''}(\tau)
+2\frac{a'(\tau)}{a(\tau)}h^{ (\sigma) }_k {'}(\tau)
+k^2 h^{(\sigma)}_k(\tau )=0,
\ee
where a prime means taking derivative with respect to  $\tau$.
The two polarizations of
$h^{(\sigma)}_k(\tau )$ have the same statistical properties and give equal contributions to the unpolarized RGWs background,
so the super index $(\sigma)$ can be dropped. The approximate solutions of Eq. (\ref{eq})
are well analyzed in \cite{grishchuk,grishchuk3,zhang2}, and are detailed listed  in \cite{Tong6} given an accelerating universe at present. Furthermore, the analytic solutions were also studied by many authors \cite{Zhang4,Yuki,Miao,Kuroyanagi,TongZhang}.  For a power-law form of
the scale factor
$a(\tau) \propto \tau^\alpha$,
the analytic solution to Eq.(\ref{eq}) is a linear combination of
Bessel and Neumann functions
\be \label{hom}
h_k(\tau)=\tau^{\frac{1}{2}-\alpha}
 \big[C_1 J_{\alpha-\frac{1}{2}}(k \tau)
      +C_2   N_{\alpha-\frac{1}{2}}(k \tau)\big],
\ee
where the constants $C_1$ and $C_2$ for each stage are determined
by the continuities of $h_k(\tau)$ and  $h'_k(\tau)$
at the joining points
$\tau_1,\tau_s,\tau_2$ and $\tau_E$ \cite{Zhang4,Miao}.
Therefore, the all the constants in the solutions  of RGWs can be completely fixed,
once the initial condition  is given.
In  a spatially flat ($k=0$)  universe, the scale factor indeed has a power-law form in various stages \cite{grishchuk,Miao,TongZhang,Tong6}.
It is described by the following successive stages
:

The inflationary stage:
\be \label{inflation}
a(\tau)=l_0|\tau|^{1+\beta},\,\,\,\,-\infty<\tau\leq \tau_1,
\ee
where the inflation index $\beta$ is an  model parameter describing the expansion history during inflation.
The special case of $\beta=-2$  corresponds the exact de Sitter expansion.
However, both the model-predicted and the observed results
indicate that the value of $\beta$ could differ slightly from  $-2$.

The preheating stage :
\be\label{betas}
a(\tau)=a_z|\tau-\tau_p|^{1+\beta_s},\,\,\,\,\tau_1\leq \tau\leq \tau_s,
\ee
where the parameter  $\beta_s$ describes the expansion behavior of the preheating stage from the end of inflation to the happening of reheating process followed by the radiation-dominant stage. In some literatures \cite{Starobinsky2,Tong6}, $\beta_s$ is  set to be $1$ , however,  we take $\beta_s$ as a free parameter in the paper.

The radiation-dominant stage :
\be \label{r}
a(\tau)=a_e(\tau-\tau_e),\,\,\,\,\tau_s\leq \tau\leq \tau_2.
\ee

The matter-dominant stage:
\be \label{m}
a(\tau)=a_m(\tau-\tau_m)^2,\,\,\,\,\tau_2 \leq \tau\leq \tau_E.
\ee

The accelerating stage up to the present time $\tau_0$
\cite{zhang2}:
\be \label{accel}
a(\tau)=l_H|\tau-\tau_a|^{-\gamma},\,\,\,\,\tau_E \leq \tau\leq
\tau_0,
 \ee
 where $\gamma$ is a $\Omega_\Lambda$-dependent
parameter, and $\Omega_\Lambda$ is the energy density contrast.
To be specific, we take $\gamma\simeq 1.97$
\cite{zhangtong} for $\Omega_{\Lambda}=0.73 $ \cite{Komatsu} in this paper.
It is convenient to choose the normalization  $|\tau_0-\tau_a|=1$, i.e.,
the present scale factor $a(\tau_0)=l_H$. From the definition of the
Hubble constant, one has $l_H=\gamma/H_0$,
where $H_0=100\, h$ km s$^{-1}$Mpc$^{-1}$ is the present Hubble constant.
We  take  $h\simeq 0.704$ \cite{Komatsu} throughout this paper.
Supposing  $\beta$ and $\beta_s$ are  model parameters, all the constants included trough Eq.(\ref{inflation}) to Eq. (\ref{accel}) can be fixed by the continuity of $a(\tau)$ and $a'(\tau)$ at the four given
joining points $\tau_1$, $\tau_s$, $\tau_2$ and $\tau_E$, if  one  knows the increases of the scale factor  of various stages, i.e., the definite values of
$\zeta_1\equiv{a(\tau_s)}/{a(\tau_1)}$,
$\zeta_s\equiv{a(\tau_2)}/{a(\tau_s)}$,
$\zeta_2\equiv{a(\tau_E)}/{a(\tau_2)}$,
and
$\zeta_E\equiv{a(\tau_0)}/{a(\tau_E)}$.

The spectrum of RGWs $h(k,\tau)$ is defined by
\be \langle
h^{ij}(\tau,\mathbf{x})h_{ij}(\tau,\mathbf{x})\rangle\equiv\int_0^\infty
h^2(k,\tau)\frac{dk}{k},
\ee
where the angle brackets mean ensemble
average. The dimensionless spectrum $h(k,\tau)$ relates to the mode $h_k(\tau)$ as
\cite{TongZhang} \be \label{relation0}
h(k,\tau)=\frac{\sqrt{2}}{\pi}k^{3/2} |h_k(\tau)|.
\ee
The one that we are of interest is  the present RGWs spectrum $h(k,\tau_0)$.
 The characteristic comoving wave number at a certain joining time $\tau_x$ is give by \cite{Tong6}
 \be\label{wavenumber}
k_x\equiv k(\tau_x) = \frac{2\pi a(\tau_x)}{1/H(\tau_x)}.
\ee
After a long but simple calculation, it is easily to obtain $k_H= 2\pi \gamma$ and the following relations:
\be\label{frelation}
  \frac{k_E}{k_H}
    = \zeta_E^{-\frac{1}{\gamma}},\quad
    \frac{k_2}{k_E}=\zeta_2^{\frac{1}{2}},\quad
    \frac{k_s}{k_2}=\zeta_s, \quad
    \frac{k_1}{k_s}=\zeta_1^{\frac{1}{1+\beta_s}}.
 \ee
In the present universe,
the physical frequency relates to a comoving wave number  $k$ as
\be \label{freq}
f=  \frac{k}{2\pi a (\tau_0)} = \frac{k}{2\pi l_H}.
\ee
 The present energy density contrast of RGWs defined by
$ \Omega_{GW}=\langle\rho_{g}\rangle/{\rho_c}$,
where $\rho_g=\frac{1}{32\pi G}h_{ij,0}h^{ij}_{,0}$
is the energy density of RGWs
and $\rho_c=3H_0^2/8\pi G$ is the critical energy density, is given by
 \cite{grishchuk3,Maggiore}
\be\label{gwe}
\Omega_{GW}=
\int_{f_{low}}^{f_{upper}} \Omega_{g}(f)\frac{df}{f},
\ee
with
\be\label{omega}
\Omega_{g}(f)=\frac{2\pi^2}{3}
        h^2_c(f)
     \Big(\frac{f}{H_0}\Big)^2
\ee
being the dimensionless  energy density spectrum. We have used  a new notation, $h_c(f)=h(f,\tau_0)/\sqrt{2}$,
called {\it characteristic strain spectrum} \cite{Maggiore} or {\it chirp amplitude} \cite{Boyle}.
The lower and upper limit of integration in Eq.(\ref{gwe})
can be taken to be  $f_{low}\simeq f_E$
and  $f_{upper}\simeq f_1$, respectively, since only the wavelength of the modes inside the horizon contribute to the total energy density.

\section{  The increases of the scale factor }

 For the simple $\Lambda$CDM model, the late-time acceleration of the universe is well know.
One easily has
$\zeta_E =1+z_E=({\Omega_\Lambda}/{\Omega_m})^{1/3}\simeq1.4$, where $z_E$ is
the redshift when the  accelerating expansion begins.
The increase of  the scale factor  duration of the matter-dominated
stage can also be obtained straightforwardly, $\zeta_2
=\frac{a(\tau_0)}{a(\tau_2)} \frac{a(\tau_E)}{a(\tau_0)}
=(1+z_{eq}) \zeta_E^{-1}$ with $z_{eq}=3240$ \cite{Komatsu}.
However, the histories of the radiation-dominated stage and the preheating stage are
not known well. Recently, Mielczarek \cite{Mielczarek} proposed a method to
evaluate the reheating temperature, $T_{\rm{RH}}$, under the frame of the slow-roll inflation model with a quadratic potential $V(\phi)=\frac{1}{2}m^2\phi^2$ combing the observations
from WMAP. Using this method, $\zeta_s$ and $\zeta_1$ can be determined subsequently with the evaluation of $\trh$ \cite{Tong6}. In this paper, we consider a more general power-law form of the potential, $V(\phi)=\lambda\phi^n$, where $\lambda$ is a constant. For this general form of $V(\phi)$, it is hard to obtain the analytic expression of the energy density of the universe at the end of inflation, and in turn, it is hard to obtain the temperature of reheating analytically. Hence, we will take some reasonable values of $\trh$ constrained by CMB observations\cite{martin}.

Firstly, we discuss the value of  $\zeta_s$.  After  reheating,
the universe is filled with the relativistic plasma, which undergoes a adiabatic  expansion  as long as the entropy transfer between the radiation and other components can be neglected.
The adiabatic approximation leads to the  conservation of the entropy, i.e., $dS=0$. It  implies $sa^3=$const,
where the entropy density $s$ of radiation is given by
  \be
  s=\frac{2\pi^2}{45}g_s T^3.
  \ee
 Here, $g_s$  counts  the effective number of relativistic species contributing to the radiation entropy.
Another similar  quantity  $g$, counting the effective number of relativistic species contributing to the energy density of radiation,  relates to energy density:
 \be
 \rho=\frac{\pi^2}{30}gT^4.
 \ee
 The behavior of $g$ and $g_s$ with different energy scale were demonstrated in \cite{Yuki}. At the energy above $\sim0.1$ MeV, one has $g=g_s$. Moreover, at the energy scales above $\sim1$ TeV, $g=106.75$ in the standard model, and  $g\simeq220$ in the minimal extension of supersymmetric standard model, respectively. On the other hand, at the energy scales below $\sim0.1$ MeV, $g=3.36$ and $g_s=3.91$ respectively.
 According to the conservation of entropy, one can easily gets  the increase of the scale factor from the reheating till  the recombination \cite{Mielczarek},
 \be\label{delta1}
\frac{a_{rec}}{a(\tau_s)}=\frac{T_{\rm{RH}}}{T_{rec}}\left(\frac{g_{\ast s}}{g_{\star s}}\right)^{1/3},
\ee
where $a_{rec}$  and $T_{rec}$ stand  for the scale factor and the temperature at the recombination, respectively.
 $g_{\ast s}$ and  $g_{\star s}$ count  the effective number of relativistic species contributing to the entropy during the reheating and that during recombination, respectively. As discussed in \cite{martin}, the lower band of the reheating energy scale is $17.3$ TeV constrained by the observed scalar power spectrum of CMB at $95\%$ of the confidence limit. Thus,   in this paper we assume $g_{\ast s}\simeq200$ eclectically, which was also employed in \cite{martin}. On the other hand, one has  $g_{\star s}=3.91$ including the contributions of the effective number from  photons and  three species of massless neutrinos to the radiation entropy  during the recombination, since  the energy scale at the recombination  $T_{rec}=T_{\rm{CMB}}(1+z_{rec})\sim 10^{-7}$ MeV. Under the assumption of $g_{\ast s}\simeq200$, the lower bound of $\trh\gtrsim 6\cdot10^3$ GeV was obtained \cite{martin}. On the other hand, gravitinos production gives an upper bound \cite{gravitinos}. For instance, in the framework of the Constrained minimal supersymmetric standard model \cite{Bailly},  the upper bound of $\trh$  was found that $\trh\lesssim$ a few $\times10^7$ GeV from over-production of $^6\rm{Li}$ from bound state effects, and moreover, $\trh$ can be relaxed to $\lesssim$ a few $\times10^8$ GeV  when a more conservative bound on $^6\rm{Li}/^7\rm{Li}$ was used. However, if one does not consider the gravitinos production problem, the most upper bound of $\trh$ could be up to  $\lesssim3\cdot10^{15}$ GeV coming from the energy scale at the end of inflation \cite{martin}.  Based on Eq. (\ref{delta1}), one easily obtain
\be\label{zetas}
\zeta_s=\frac{a(\tau_2)}{a_{rec}}\frac{a_{rec}}{a(\tau_s)}=\frac{T_{\rm{RH}}}{T_{\rm{CMB}}(1+z_{eq})}\left(\frac{g_{\ast s}}{g_{\star s}}\right)^{1/3},
\ee
where we have used $T_{rec}=T_{\rm{CMB}}(1+z_{rec})$. With  $T_{\rm{CMB}}=2.725\ {\rm{K}}=2.348\cdot10^{-13}$ GeV \cite{Komatsu}, one has $\zeta_s\simeq5\times10^{16}$ for example.
Secondly, we discuss the evaluation of $\zeta_1$. First of all, we briefly recall the slow-roll inflation model.
For  slow-roll inflation,
the evolution is described by the usual  slow-roll parameters \cite{LiddleA}:
\be\label{slow}
\epsilon\equiv\frac{m^2_{Pl}}{16\pi}\left(\frac{V'}{V}\right)^2, \qquad
\eta\equiv\frac{m^2_{Pl}}{8\pi}\frac{V''}{V}, 
\ee
which are required to be much small than unity  for the slow-roll approximation to be valid. $\epsilon$ approaches  to unity at the end of inflation. When the slow-roll conditions are satisfied, inflation continues keeping the Hubble rate nearly constant, and
 the primordial tensor power spectrum and the
scalar power spectrum are respectively given as \cite{Boyle,Kuroyanagi}:
\ba\label{tensor}
&&\Delta_h^2(k,\tau_\ast)\approx\frac{16}{\pi}\left(\frac{H_\ast}{m_{\rm{Pl}}}\right)^2,\\
&&\Delta_\mathcal{R}^2(k,\tau_\ast)\approx\frac{1}{\pi\epsilon}\left(\frac{H_\ast}{m_{\rm{Pl}}}\right)^2,\label{scalar}
\ea
where $H_\ast$ is the   Hubble rate during  inflation, and
$\tau_\ast$ stands for the moment when the $k$-mode exits the horizon.
On the other hand, based on the observations of CMB, the present scalar power spectrum can be expanded in power laws,
\ba \label{tensor2}
&&\Delta_h^2(k)=\Delta_h^2(k_0)\left(\frac{k}{k_0}\right)^{n_t},\\
&&\Delta_\mathcal{R}^2(k)=\Delta_\mathcal{R}^2(k_0)\left(\frac{k}{k_0}\right)^{n_s-1}\label{scalar2},
\ea
where $\Delta^2_h(k_0)$ and $\Delta^2_{\mathcal{R}}(k_0)$ are the power spectrum of the tensor perturbations and
curvature perturbations evaluated at the pivot wave number $\k0=k_0/a(\tau_0)=0.002$ Mpc$^{-1}$ \cite{Komatsu}, respectively. Furthermore, under the slow-roll approximation, at the pivot wave number $k_0$  the spectral parameters are given by \cite{LiddleA}
\ba\label{nt}
&&n_t\simeq-2\epsilon,\\ \label{ns}
&&n_s\simeq1-6\epsilon+2\eta,
\ea
In general, the  spectral indices  $n_t$ and $n_s$ are
   $k-$depedent, described by the running parameters $\alpha_t\equiv dn_t/d \ln k$ and $\alpha_s\equiv dn_s/d \ln k$, respectively \cite{grishchuk91,Kosowsky,LiddleA,TongZhang}.  However, $\alpha_t$ and $\alpha_s$ are only second order small quantities. Moreover, if one uses the quantum normalization (see below) as the initial condition for    the generation of RGWs, $\alpha_t$ should  be exactly zero. On the other hand, as will be seen below,  non-zero $\alpha_s$ would induce an $n_s$ greater than 1, which make us difficult to evaluate the increase of the scale factor from the $k_0$ mode exiting the horizon during inflation to the end of inflation.
    Hence, in this paper we will simply  set $\alpha_t=\alpha_s=0$.
 Note that  Even though
the value  of  $n_t$ is quite uncertain, $n_s$ can be well  constrained by CMB \cite{Komatsu} or BAO \cite{Sanchez}.
The ratio of the primordial tensor power spectrum to the  scalar power spectrum is defined as \cite{Boyle,Kuroyanagi}
\be\label{ratio2}
r\equiv \frac{\Delta_h^2(k,\tau_\ast)}{\Delta_\mathcal{R}^2(k,\tau_\ast)}=16\epsilon,
\ee
based on  Eqs.(\ref{tensor}) and (\ref{scalar}). Therefore, at the pivot  number $k_0$, one has
\be\label{r2}
r=\frac{\Delta_h^2(k_0,\tau_i)}{\Delta_\mathcal{R}^2(k_0,\tau_i)}\simeq \frac{\Delta_h^2(k_0)}{\Delta_\mathcal{R}^2(k_0)},
\ee
where $\tau_i$ is the $k_0$-mode exit the horizon during inflation. The approximation of the second equation in Eq.(\ref{r2}) accounts for that the pivot $k_0$ wave mode reentered the horizon a little earlier than the present time, and then has suffered a decay. Therefore, the ratio ${\Delta_h^2(k_0)}/{\Delta_\mathcal{R}^2(k_0)}$ can not exactly reflect the true value of $r$ given by its definition, however, the deviation would be expected to be less than $\sim0.8\%$ \cite{Tong6}. Hence, we will use this approximation when confront with the CMB observations. Furthermore, under this approximation, one has a simple relation:
 \be\label{nt2}
 n_t=2\beta+4,
 \ee
 since the primordial spectrum of RGWs has a power-law form $\Delta(k_0)\simeq\Delta(k_0,\tau_i)\propto k^{2\beta+4}$ \cite{Tong6}.
  WMAP\,7 Mean \cite{Komatsu} fixed $\Delta^2_{\mathcal{R}}(k_0)\equiv A_s=(2.43\pm 0.11)\cdot10^{-9}$. Thus, the non-zero value of $r$ implies the existence of gravitational wave background, which induced uniquely the B-mode polarization of CMB \cite{Amarie}. At present only
observational constraints on $r$ have been given
\cite{Komatsu,Hinshaw}. The upper bounds of $r$ are
recently constrained \cite{Komatsu}  as  $r<0.24$ by WMAP+BAO+$H_0$ and $r<0.36$ by
WMAP 7 only for  $\alpha_s=0$, and  $r<0.49$  for  $\alpha_s\neq0$
by both the combination of WMAP+BAO+$H_0$ and the WMAP 7 only,
respectively. Furthermore, using a discrete, model-independent measure of the degree
of fine-tuning required, if $0.95\lesssim n_s<0.98$, in accord with current measurements,
 the tensor-to-ratio satisfies $r\gtrsim10^{-2}$  \cite{Boyle2}.
Therefore, one can normalize the RGWs at $k=k_0$ using Eq. (\ref{r2}), if $r$ can be determined definitely.

As analyzed by Mielczarek \cite{Mielczarek}, for the pivot wave number $\k0$, the total increase of the scale factor from the mode exit the horizon during inflation up to the present time can be evaluated as
\be\label{atot}
\zeta_{\rm{tot}}\simeq \frac{H_\ast}{\k0}.
\ee
Due to Eqs. (\ref{scalar}) and (\ref{scalar2}), one has
\be\label{k0scalar}
\frac{1}{\pi\epsilon}\left(\frac{H_\ast}{m_{\rm{Pl}}}\right)^2\approx \Delta_\mathcal{R}^2(k_0),
\ee
where  the approximation $\Delta_\mathcal{R}^2(k_0)\approx\Delta_\mathcal{R}^2(k_0,\tau_i)$
was used. Taking the form $V(\phi)=\lambda\phi^n$,  one can easily have a relation:
\be\label{epsilon}
\epsilon=\frac{n(1-n_s)}{2(n+2)}
\ee
from  Eqs. (\ref{slow}) and (\ref{ns}).
Plugging Eqs. (\ref{k0scalar}) and (\ref{epsilon}) into  Eq. (\ref{atot}) gives
\be
\zeta_{\rm{tot}}\simeq \frac{m_{\rm{Pl}}}{\k0}\sqrt{\frac{\pi n}{2(n+2)}(1-n_s)\Delta_\mathcal{R}^2(k_0)}.
\ee
On the other hand, if we assume the universe did a quasi-de Sitter expansion($\beta\approx-2$), the increase  of a scalar factor from the moment of $k_0$ mode exiting the horizon during inflation to the end of inflation is give by
\be\label{zetai}
\zeta_i=e^{N},
\ee
where $N$ is the e-folding number, which can be estimated as
\be
N\simeq-\frac{8\pi}{m^2_{\rm{Pl}}}\int_{\phi_{obs}}^0\frac{V(\phi)}{V'(\phi)}d\phi.
\ee
Concretely, for $V(\phi)=\lambda\phi^n$,   one can get
\be\label{nnumber}
N\simeq\frac{n+2}{2(1-n_s)},
\ee
with the help of  Eqs. (\ref{slow}) and (\ref{ns}).
So, if $n=2$, Eq.(\ref{nnumber}) reduces to the result shown in \cite{Mielczarek}.
Plugging Eq. (\ref{nnumber}) into Eq. (\ref{zetai}), and using the identity
\be
\zeta_{\rm{tot}}=\zeta_i\zeta_1\zeta_s\zeta_2\zeta_E,
\ee
one can easily obtain   the complete expression of $\zeta_1$:
\be\label{zeta1}
\zeta_1=\frac{m_{\rm{Pl}}}{\k0}\left[\pi \Delta_\mathcal{R}^2(k_0)(1-n_s)\frac{n}{2(n+2)}\right]^{1/2}
\frac{T_{\rm{CMB}}}{T_{\rm{RH}}}\left(\frac{g_{\star s}}{g_{\ast s}}\right)^{1/3}\exp{\left[-\frac{n+2}{2(1-n_s)}\right]}.
\ee
One can examine that, for $n=2$, the above expression reduces to Eq. (11) in Ref.\cite{Tong6} after using Eq.(7) in the same reference.
In the following, let us see the reasonable range of the index $n$ constrained by both theories and observations.
As well known, at the end of inflation, the scalar field $\phi$ oscillates quickly around some point where $V(\phi)$ has a minimum. In the limit that the oscillation rate is much greater than Hubble expansion rate $H$, and ignoring the coupling between the scalar field $\phi$ and other components, it is found that \cite{Turner} the scalar field oscillations behave like a fluid with $p=\bar{w}\rho$, where the average equation of state $\bar{w}$ depends on the form of the potential $V(\phi)$. For $V(\phi)=\lambda\phi^n$, one has
 \be\label{eos}
\bar{w}=\frac{n-2}{n+2}
\ee
 and
$\rho$ decreases as $a^{-6n/(n+2)}$. In particular, $n=2$, one has $\bar{w}=0$ and $\rho\propto a^{-3} $, which imply a matter-dominant like expansion of the preheating stage \cite{Starobinsky2}. Adding the consideration of the coupling between the scalar field and the resulting relativistic particle creation, Martin and Ringeval \cite{martin} verified the relation (\ref{eos}) using a numerical method, and it was found that the average $\bar{w}$ never deviates from zero exceeding $8\%$. From theoretical consideration, one should have $\bar{w}<1$ to satisfy the positivity energy conditions; while $\bar{w}>-1/3$ to make sure the inflation must stop and the preheating stage begins. Due to Eq. (\ref{eos}), the condition $-1/3<\bar{w}<1$ leads to $n>1$. On the other hand,
Martin and Ringeval \cite{martin} firstly gave a constraint on $n$ based on the CMB observations, $n<2.1$.
Therefore, based on both the theories and observations, the index $n$ is constrained to be
\be\label{nrange}
1<n<2.1.
\ee
Note that, there is a relation between $n$ and $\beta_s$. According to the energy conservation equation and the
Friedmann Equation,
\ba
&&\dot{\rho}+3\frac{\dot{a}}{a}\rho(1+\bar{w})=0,\\
&&\left(\frac{\dot{a}}{a}\right)^2=\frac{8\pi G}{3}\rho,
\ea
one can easily obtain $a\propto t^{2/(3+3\bar{w})}\propto\tau^{{2}/{(1+3\bar{w})}}$. Using Eqs. (\ref{betas}) and (\ref{eos}), and allowing for $\rho\propto a^{-6n/(n+2)}$, one has
\be\label{betasn}
\beta_s=\frac{4-n}{2(n-1)}.
\ee
Then, in principle,  the expression of $\zeta_1$ in Eq. (\ref{zeta1}) can be rewritten as a function of $\beta_s$.
From the  combination of Eqs. (\ref{nrange}) and (\ref{betasn}), one finds that, $n>1$ leads to $\beta_s>-0.5$ and
$n<2.1$ leads to $\beta_s>0.86$, respectively. Based on the  range of $n$ (or $\beta_s$) discussed above, we try to constrain some parameters combining with CMB observations.

\section{ Parameters constraints from observations}

As shown in the previous section, many parameters are dependent on the value of $n_s$. Seven-year WMAP Mean \cite{Komatsu} gives
$n_s=0.967\pm0.014$, and $n_s=0.982^{+0.020}_{-0.019}$ when one also considers the tensor mode  contributions to the anisotropies of CMB. Moreover, the combination WMAP+BAO+$H_0$ Mean gives $n_s=0.968\pm0.012$, and $n_s=0.973\pm0.014$ when  the tensor mode  contributions are included.
 Independently, SDSS III predicts $n_s=0.96\pm0.009$ \cite{Sanchez}. As can be seen in Eq. (\ref{zeta1}),  $\zeta$ is sensitively dependent on $n_s$, and in turn one can expect that the spectrum of RGWs  also depend sensitively on $n_s$ in the very high frequencies. Therefore, for a general demonstration,  we consider the cases: $n_s=0.96, 0.97$ and $0.98$, respectively.

  \begin{figure}
\resizebox{90mm}{!}{\includegraphics{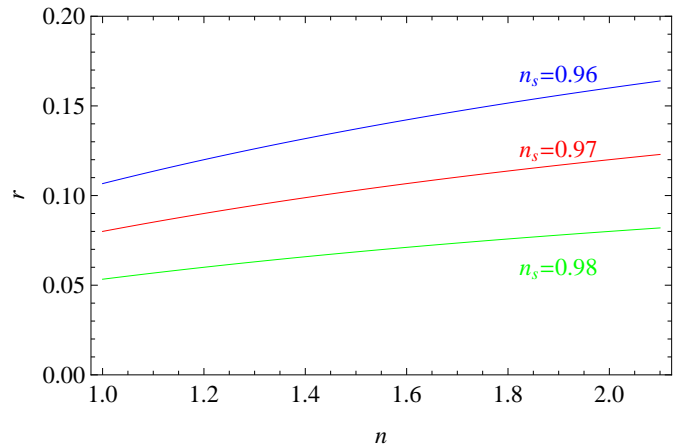}
}
\caption{\label{rn}
The relation between $r$ and $n$ for the fixe value of $n_s=0.96$,  $n_s=0.97$ and $n_s=0.98$, respectively.
}
\end{figure}
     \begin{figure}
\resizebox{90mm}{!}{\includegraphics{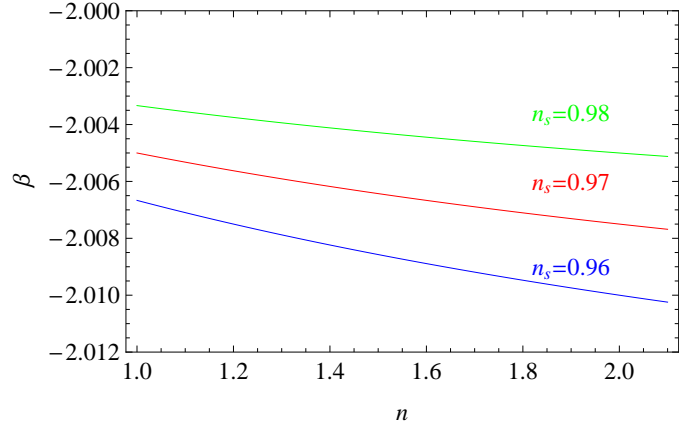}
}
\caption{\label{betan}
The relation between $\beta$ and $n$ for the fixe value of $n_s=0.96$,  $n_s=0.97$ and $n_s=0.98$, respectively.
}
\end{figure}
  Firstly, let us constrain the tensor-to-scalar ratio $r$. According to Eqs. (\ref{slow}) and (\ref{ratio2}),
  it is straightly to get
  \be\label{r3}
  r=\frac{8n}{n+2}(1-n_s).
  \ee
  We show this relation in Fig.\ref{rn}. One can see that $r$ increases slowly with $n$.
  $r$ lies in (0.11, 0.16), (0.08, 0.12), and (0.05, 0.08) for $n_s=0.96$, $n_s=0.97$, and $n_s=0.98$, respectively. Similarly, from Eqs. (\ref{nt}) and (\ref{nt2}), one has
  \be\label{beta3}
  \beta=-2-\frac{n}{2(n+2)}(1-n_s),
  \ee
  which is shown in Fig.\ref{betan}. The parameter $\beta$ is constrained in the range of $(-2.007,-2.010)$, $(-2.005,-2.008)$, and $(-2.003, -2.005)$ for $n_s=0.96$, $n_s=0.97$, and $n_s=0.98$, respectively.
  Therefore, the  range of $n$ in Eq. (\ref{nrange}) leads to very tight constraints on $r$ and $\beta$, which are
  limited in very narrow ranges with definite value of $n_s$.

\begin{figure}
\resizebox{110mm}{!}{\includegraphics{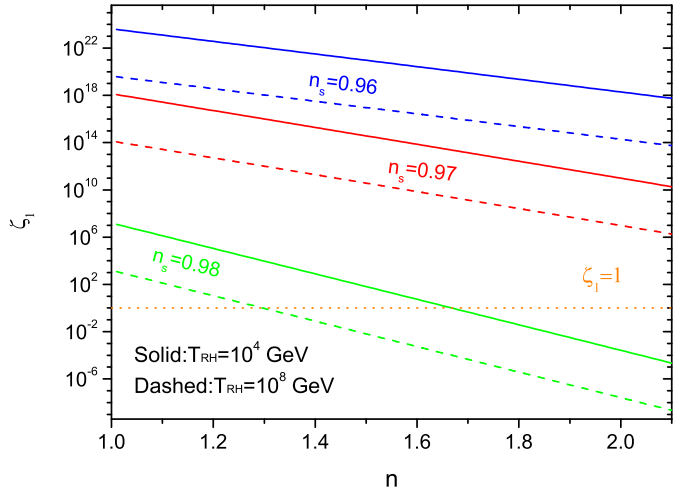}
}
\caption{\label{zeta1plot}
$\zeta_1$ as a function of $n$ for the fixed value of $n_s=0.96$,  $n_s=0.97$ and $n_s=0.98$, respectively.
The solid lines and dashed lines   correspond  to $\trh=10^4$ GeV and $\trh=10^8$ GeV, respectively. The dotted line represents $\zeta_1=1$.
}
\end{figure}

  Now, let us see the increase of the scale factor during preheating stage $\zeta_1$, which is expressed in Eq.(\ref{zeta1}). We plot it in Fig.\ref{zeta1plot} as a function of $n$ with definite values of $\trh$.
Allowing for the expansion of the universe, one would expect that $\zeta_1>1$. As can be seen in Fig.\ref{zeta1plot},  the cases of $n_s=0.96$ and $0.97$ can make sure well  the resultant $\zeta_1$ being much
larger than 1, however, the case of $n_s=0.98$ can not be compatible with the fact $\zeta_1>1$ in the whole range of $n$ shown in Eq.(\ref{nrange}). If $n_s$ is determined well to be as high as $0.98$, it will give  very tight
constraints on $n$. Concretely, $n\lesssim1.7$ and $n\lesssim1.3$ for $\trh=10^4$ GeV and  $\trh=10^8$ GeV, respectively.
  \begin{figure}
\resizebox{110mm}{!}{\includegraphics{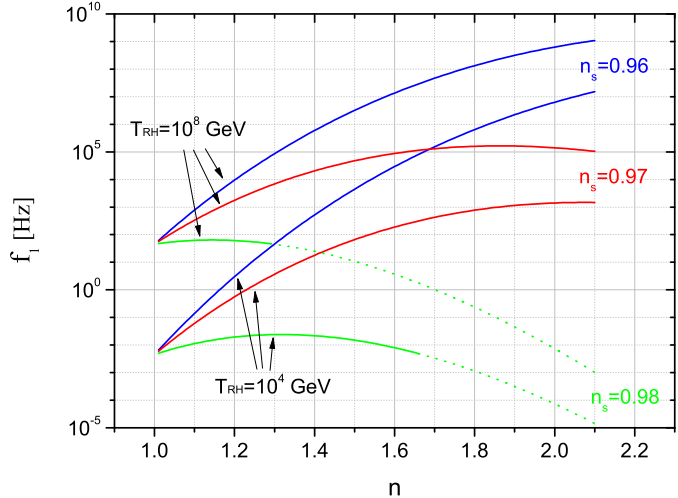}
}
\caption{\label{f1plot}
The upper limit frequency $f_1$ as a function of $n$ for different combinations of  $n_s$ and $\trh$. The dotted parts
for $n_s=0.98$ are constrained by the condition $\zeta_1>1$ shown in Fig. \ref{zeta1plot}.
}
\end{figure}
What we are more interesting are the characteristic frequencies given by Eq.(\ref{freq}). With the help of Eq. (\ref{frelation}), one can easily get the  characteristic   frequencies: $f_H=H_0\simeq 2.28\cdot10^{-18}$ Hz,
  $f_H=H_0\simeq 2.28\cdot10^{-18}$ Hz,
 $f_E\simeq1.93\cdot10^{-18}$ Hz,
 $f_2\simeq 9.3\cdot10^{-17}$ Hz.
  The value of $f_s$ depends linearly on $\trh$. For instance,
 $f_s\simeq 4.54\cdot10^{-3}$ Hz for $\trh=10^4$ GeV and
 $f_s\simeq 45.4$ Hz for $\trh=10^8$ GeV, respectively.
 Since $f_1=f_s\zeta_1^{\frac{1}{1+\beta_s}}$, it depends on the values of $n$ (or $\beta_s$), $n_s$ and $\trh$.
 It is worth to be study in detail, since the characteristic frequency $f_1$ is approximately  the highest frequency of RGWs. The modes whose frequency higher than $f_1$ decay with the expansion of the universe \cite{grishchuk,grishchuk3}.
 We plot $f_1$ as a function of $n$ with definite $n_s$  and $\trh$ in Fig.\ref{f1plot}.
 One can see that the behaviors of $f_1$ along with the increasing $n$ are quite different for different values of $n_s$. For the case of $n_s=0.98$, we plotted $f_1$ using dotted lines for the values of $n$ constrained by
 $\zeta_1>1$ which are shown in Fig.\ref{zeta1plot}. In the part of large  $n$, $f_1$ is larger for smaller values of $n_s$. On the other hand,
 in the limit of $n\rightarrow1$, $f_1$ becomes a  fixed value independent on $n_s$, and moreover, the asymptotic fixed $f_1$ is larger for a larger value of $\trh$. It is easy to understand if one has found that $\beta_s\rightarrow+\infty$ as  $n\rightarrow1$ from Eq.(\ref{betasn})  which leads to $f_1\rightarrow f_s$.
   The value  of $f_1$ should be below the  constraint  from the rate of
 the primordial nucleosynthesis, $f_1\lesssim3\times10^{10}$ Hz
 \cite{grishchuk}. When the acceleration epoch is considered, the constraint becomes
  $f_1\simeq4\times10^{10}$ Hz. This will in turn give some constraints on $n$, $n_s$ and $\trh$.

\begin{figure}
\resizebox{110mm}{!}{\includegraphics{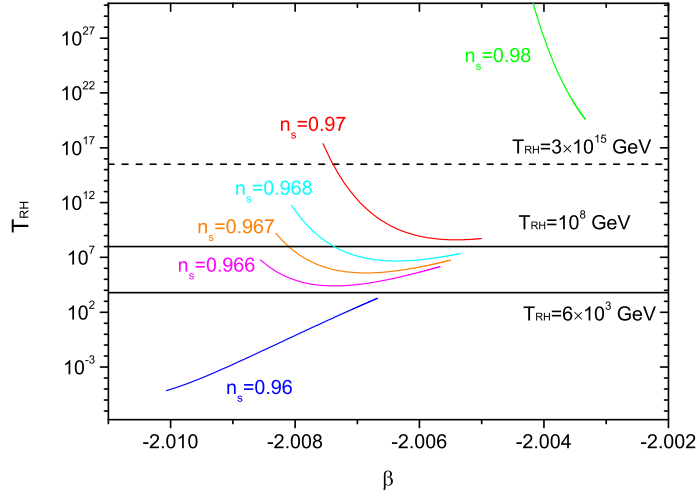}
}
\caption{\label{trhbeta}
The relation between $\trh$ and $\beta$ based on Eq.(\ref{arelation}) due to the condition of quantum normalization. }
\end{figure}

As analyzed in our previous work \cite{Tong6}, when  the quantum normalization for the generation of RGWs during inflation is employed, one has
\be\label{arelation}
\Delta_\mathcal{R}(k_0)r^{1/2}=8\sqrt{\pi}l_{Pl}H_0
\zeta_1^{\frac{\beta_s-\beta}{1+\beta_s}}\zeta_s^{-\beta} \zeta_2^{\frac{1-\beta}{2}}
  \zeta_E^{\frac{\beta-1}{\gamma}}\left(\frac{k_0}{k_H}\right)^\beta,
\ee
where $l_{Pl}=\sqrt{G}$ is the Planck length. In Eq.(\ref{arelation}), there are totally six parameters: $r$, $\beta$, $\beta_s$, $n$, $\trh$ and $n_s$. However, among them  only three  are independent, due to Eqs.(\ref{betasn}), (\ref{r3}) and (\ref{beta3}). We show the $\trh-\beta$ relation with definite values of $n_s$ in Fig.\ref{trhbeta}. First of all, we define the range of $6\cdot10^3-10^{8}$ GeV as Region I; while the
 range of $6\cdot10^3-3\cdot10^{15}$ GeV as Region II. It is found that, under the condition of quantum normalization, $n_s=0.96$ and $n_s=0.98$ can
 be ruled out, since the resultant $\trh$  outside  Region II. If one consider the gravitinos production problem, the case $n_s=0.97$ would also be ruled out, since  the resultant $\trh$  outside  Region I. However, the resultant $\trh$ given by
$n_s=0.966$ lies well inside Region I for the whole range of $\beta$ given by Eq. (\ref{beta3}). Moreover, as shown in Fig.\ref{trhbeta}, the quantum normalization will give a little tighter constraints on the range of $\beta$ for $n_s=0.967$ and $n_s=0.968$.  Note that, these results are based on the validity  of quantum normalization, however, it is not the unique initial condition.
Let us make a comparison with the previous results in \cite{Tong6}.  Taking $n_s=0.966$ for example,   $n=2$ leads to $\trh\simeq3.4\cdot10^6$ GeV; while $\trh\simeq2.8\cdot10^{12}$ GeV shown
in Fig.1 in Ref.\cite{Tong6}. Hence, the discrepancy of $\trh$, at six orders of magnitude,  indicates that the quantum normalization may be not a good initial condition. However, one should also keep in mind that we have used many approximations, which would also contribute a lot to the discrepancy of $\trh$ discussed above.
 Note that, if one does not consider quantum normalization, the zero point energy should be removed
 or else the cosmological constant would be 120 orders of magnitude larger than observed. Some effective methods \cite{Borges} have been pointed out.
 In next section, we will demonstrate the spectra of RGWs with and without quantum normalization respectively.

\begin{figure}
\resizebox{110mm}{!}{\includegraphics{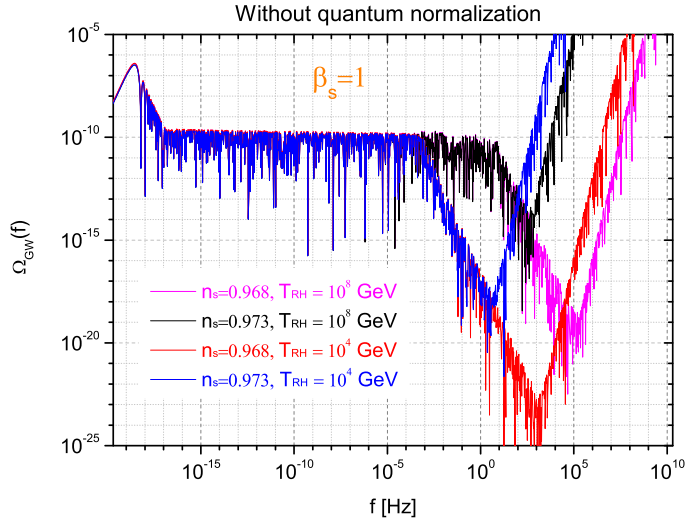}
}
\caption{\label{Powerbetas}
The energy density spectra of RGWs for the fixed $\beta_s=1$ with different combinations of $n_s$ and $\trh$, without considering the quantum normalization.
}
\end{figure}

\begin{figure}
\resizebox{110mm}{!}{\includegraphics{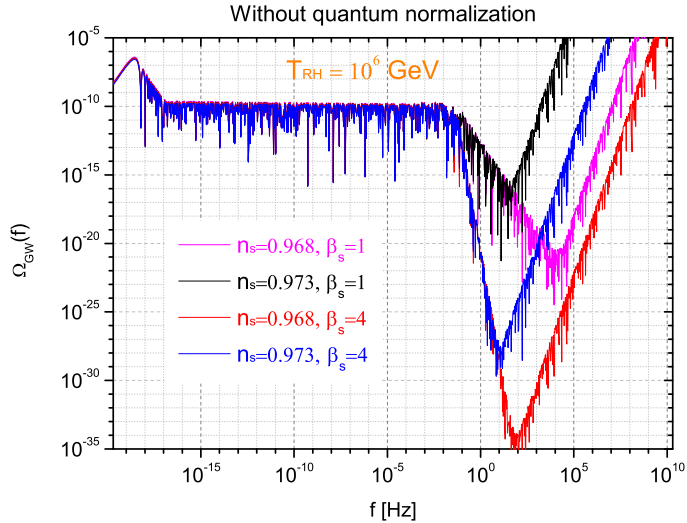}
}
\caption{\label{PowerT6}
The energy density spectra of RGWs for the fixed $\trh=10^6$ GeV with different combinations of $n_s$ and $\beta_s$, without considering the quantum normalization.
}
\end{figure}

 \section{The spectra of RGWs and their detection}

In this section, we demonstrate the energy density spectra of RGWs with reasonable values of the parameters
and discuss the detection due to the current running and planned gravitational wave detectors.

\begin{figure}
\resizebox{110mm}{!}{\includegraphics{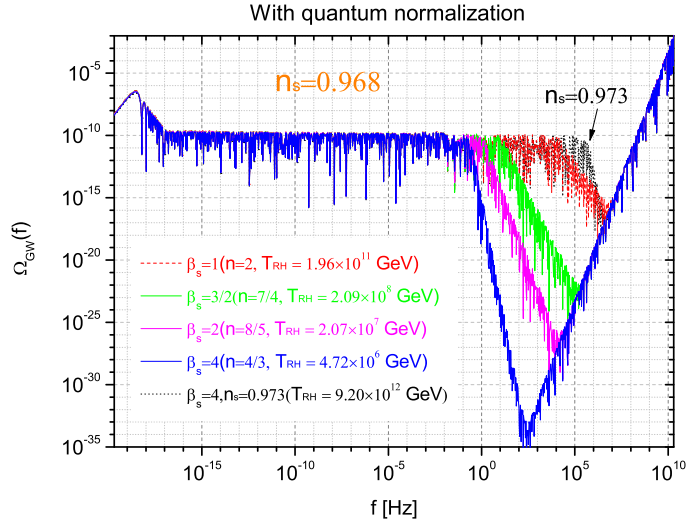}
}
\caption{\label{Powerwith}
The energy density spectra of RGWs under the condition of the quantum normalization.
}
\end{figure}

    As discussed in the previous sections, there are many parameters involved in the  spectrum of RGWs.
They are $n_s$, $n$, $r$, $\beta$, $\beta_s$, and $\trh$. However, among them only three are independent due to Eqs. (\ref{betasn})-(\ref{beta3}). Furthermore,  $n_s$ has been constrained well from observations of CMB, BAO and $H_0$. Since the spectrum of RGWs in the high frequencies extreme sensitively depends on $n_s$,  we discuss two cases of $n_s=0.968$ and $n_s=0.973$, respectively, based on the combination of WAMP+BAO+$H_0$ \cite{Komatsu}. In the following, we regard $\beta_s$ and $\trh$ as parameters, and choose some representative values of them since they have large uncertainties. In order to give a complete discussion, we will consider the spectra of RGWS both with and without quantum normalization.  As analyzed in Sec.III,   $\trh$ is constrained to be $\trh\sim10^4-10^8$ GeV, and $\beta_s$ is limited to  be  larger than $0.86$. Firstly, let us see the case of no quantum normalization.
Setting $\beta_s=1$, Fig.\ref{Powerbetas} shows the energy density spectra of RGWs, $\Omega_{\rm{GW}}(f)$, with different values of $n_s$ and $\trh$. One can see that, all the $\Omega_{\rm{GW}}(f)$  nearly  overlap each other in the low-frequencies. This is because the spectrum for $f\leq f_s$  is only related to $r$ and $\beta$ \cite{Tong6}, and, moreover, the differences of $r$ and $\beta$ are very small between the case $n_s=0.968$ and the case $n_s=0.973$ due to Eqs. (\ref{r3}) and (\ref{beta3}). Explicitly,   one has $r=0.128, \beta=-2.008$ for $n_s=0.968$   and $r=0.108, \beta=-2.007$ for $n_s=0.973$, respectively.
However, in the part of high frequencies, $\Omega_{\rm{GW}}(f)$ exhibits different properties for different combinations of $n_s$ and $\trh$. On one hand, for the same value of $\trh$, the spectrum  $\Omega_{\rm{GW}}(f)$
 with $n_s=0.968$ and that with $n_s=0.973$ have the same ``turning point'' from which $\Omega_{\rm{GW}}(f)$
decreases rapidly with the increasing frequency, and the ``turning point'' is just $f_s$ which is only dependent on $\trh$. Moreover, the decreasing slope of the logarithm of the  two spectra for $f\geq f_s$ are nearly the same since $\Omega_{\rm{GW}}(f)\propto f^{4+2\beta-2\beta_s}$ \cite{Tong6} which is reduced to $\Omega_{\rm{GW}}(f)\propto f^{-2}$
for $\beta\approx-2$ and $\beta_s=1$.
 However, the $\Omega_{\rm{GW}}(f)$ with a smaller $n_s$  has a larger upper limit frequency $f_1$ which responds to a lower amplitude of $\Omega_{\rm{GW}}(f)$. On the other hand, for the same value of $n_s$, the  $\Omega_{\rm{GW}}(f)$ with a higher $\trh$ leads to not only a larger $f_s$ but also a larger $f_1$ since $f_1\propto\trh^{1/2}$ for $\beta_s=1$ which can seen from the combination of Eqs. (\ref{frelation}), (\ref{zetas}) and (\ref{zeta1}).
 Fig.\ref{PowerT6} shows the energy density spectra of RGWs for the fixed value $\trh=10^6$ GeV.
 One can see that   a larger $\beta_s$ leads to a steeper slope of  the logarithm of $\Omega_{\rm{GW}}(f)$ and a smaller $f_1$ for the same values of $n_s$. In a word, $\trh$ determines the value of $f_s$, $\beta_s$ determines the slope of the logarithm of $\Omega_{\rm{GW}}(f)$ for the fact that $\beta\approx-2$, and $f_1$ depends on all the three parameters especially $n_s$. Secondly, let us consider the case of the quantum normalization. Due to the
 resultant Eq. (\ref{arelation}), among the three parameters $\trh$, $\beta_s$ and $n_s$ only two of them are independent. Taking $n_s$ and $\trh$ as parameters, $\Omega_{\rm{GW}}(f)$ with some combinations of the two parameters are plotted in Fig.\ref{Powerwith}.

\begin{figure}
\resizebox{100mm}{!}{\includegraphics{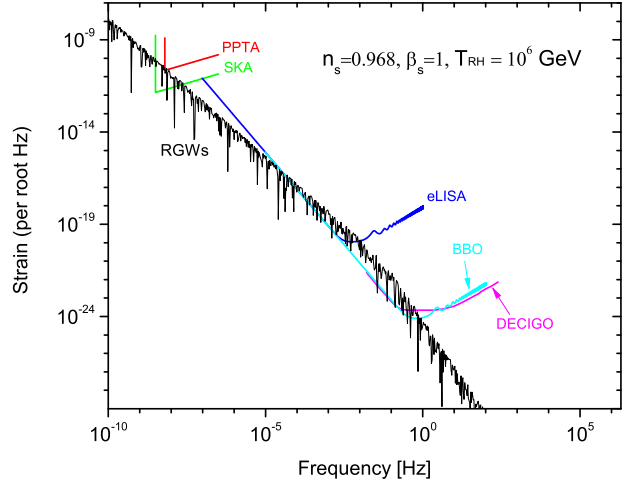}}
\caption{\label{detection}
The strain of RGWs with different parameters for $n_s=0.968$, $\beta_s=1$ and $\trh=10^6$ GeV  confronts against the current and planed GW detectors. The sensitivity curves of PPTA and SKA using pulsar timing technique are taken from Refs.\cite{Jenet} and \cite{Sesana}, respectively.
The curve of BBO is generated using the online ``Sensitivity curve generator'' \cite{lisa} with the parameters in Table II of Ref.\cite{Crowder} and Table I of Ref.\cite{Cutler}. The curve  of  DECIGO is taken from Ref.\cite{Kawamura}.}
\end{figure}

\begin{figure}
\resizebox{110mm}{!}{\includegraphics{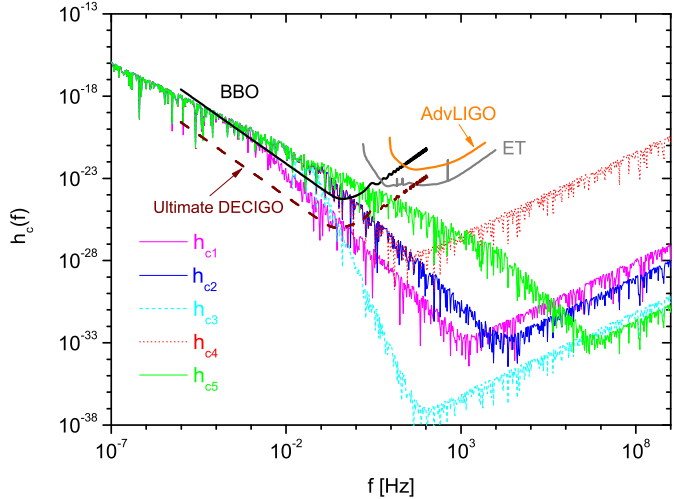}
}
\caption{\label{detection2}
The characteristic amplitude of RGWs in high frequencies confronting against the instrument noise $\sqrt{f S_n(f)}$ of BBO, Ultimate DECIGO \cite{Kudoh}, AdvLIGO and ET.
}
\end{figure}

Below, let us discuss the detection of RGWs using the ongoing and planned gravitational detectors which are sensitive at different  frequency bands. As shown in  Fig.\ref{Powerbetas}-Fig.\ref{Powerwith}, the differences of the spectra of RGWs with different parameters are only significant  in high frequency parts.
Hence, we just take a characteristic combination of the parameters $n_s=0.968$, $\beta_s=1$ and $\trh=10^6$ GeV
for demonstration.
As a conservative evaluation, in Fig. \ref{detection} we show   the strain amplitudes, $h_c(f)/\sqrt{f}$ \cite{Maggiore} of RGWs confronting  the strain sensitivity curves  of
 various gravitational wave detectors including   the complete PPTA \cite{Jenet} and SKA \cite{Sesana} using the pulsar timing technique,  and the
 space-based laser interferometers such as eLISA \cite{lisa},  BBO \cite{Crowder,Cutler}, and the Fabry-Perot DECIGO \cite{Kawamura}. One can see that, RGWs under the frame of the slow-roll inflation with a potential $V(\phi)\propto \phi^n$ are quite promising to be detected by the future  SKA, eLISA, BBO and DECIGO.
 As seen from Fig.\ref{Powerbetas}-Fig.\ref{Powerwith}, $\Omega_{\rm{GW}}(f)$ with different parameters have different properties in high frequencies. It would be interesting to discuss the detection of RGWs in high frequencies. In Fig. \ref{detection2}, we plot the characteristic amplitude of RGWs with different parameters and
 conditions compared to  the instrumental noise, $\sqrt{f S_n(f)}$, of BBO, the ultimate DECIGO \cite{Kudoh}, the  second generation ground-based laser interferometers  AdvLIGO \cite{advligo},  and  the third generation ET \cite{Hild}. The parameters and conditions of RGWs are listed in Table I. $S_n(f)$ is the normal one-side noise spectrum of detectors.
  As can be seen in Fig.\ref{detection2}, even though AdvLIGO and ET are hard to catch the signals of RGWs,
  BBO has the potential to distinguish RGWs with different parameters or different conditions in the frequency band $10^{-2}-10^0$ Hz. Furthermore, the ultimate DECIGO has the capability to distinguish
  them more easily. Thus, the future BBO and DECIGO detections
 provide an important tool not only determining
 the parameters but also examining the validity of the  quantum normalization when RGWs were generated during inflation. It is worth to point out that, at frequencies lower than $10^{-2}$ Hz the signals of RGWs are contaminated by the confusion noise produced by galactic binaries \cite{Ljz}. Hence, we should focus on the frequencies higher than $10^{-2}$ in order to distinguish various spectra of RGWs.

\begingroup
\begin{table}
\caption{\label{table}
The definitions of $h_c$ with different parameters. ``N'' stands for ``No'' meaning that the condition of the
quantum normalization is not considered; while ``Y'' stands for ``Yes'' meaning that the condition of the quantum
normalization is considered.
}
\begin{center}
\begin{tabular}{l|cccc}
\hline\hline
$\quad h_c$\quad &\quad $n_s$ &\quad $\beta_s$ &\quad $\trh$  &\quad quantum normalization\\
\hline
\quad$h_{c1}\quad$  &\quad $0.968$ & \quad $1$ &\quad $10^4$ GeV & N\\
\quad$h_{c2}\quad$ &\quad 0.968 &\quad 1 &\quad $10^6$ GeV & N\\
\quad$h_{c3}\quad$  &\quad 0.968 &\quad 4 &\quad $10^6$ GeV & N \\
\quad$h_{c4}\quad$ &\quad 0.973 &\quad 1 &\quad $10^6$ GeV & N\\
\quad$h_{c5}\quad$ &\quad 0.968 &\quad 1 &\quad $2\cdot10^{11}$ GeV & Y\\
\hline\hline
\end{tabular}
\end{center}
\end{table}
\endgroup
\section{Conclusions and Discussions}

~In the frame of the slow-roll inflation with a power-law form $V=\lambda\phi^n$, we calculated  the analytic solutions of RGWs. In the narrow range  $1<n<2.1$, the tensor-to-scalar ratio $r$ and the inflation expansion index $\beta$ are both tight limited to lie in narrow ranges for a given value of the scalar spectral index $n_s$.
Concretely, $r$ lies in (0.11, 0.16), (0.08, 0.12), and (0.05, 0.08) for $n_s=0.96$, $n_s=0.97$, and $n_s=0.98$, respectively; while $\beta$ lies in the range of $(-2.007,-2.010)$, $(-2.005,-2.008)$, and $(-2.003, -2.005)$ for $n_s=0.96$, $n_s=0.97$, and $n_s=0.98$, respectively.
Moreover, the preheating expansion index $\beta_s$ is constrained to be $\beta_s>0.86$.  We found that
the spectrum of RGWs in high frequencies  depends on the parameters $n_s$, $\beta_s$ and $\trh$. Explicitly,
$\trh$ determines $f_s$ where the flat RGWs spectrum decreases suddenly. $\beta_s$ determines  the decreasing slope of the logarithm of the spectrum. Whereas, the upper limit frequency $f_1$ is dependent on all the three parameters $n_s$, $\beta_s$ and $\trh$.  Besides, the quantum normalization for the generation of RGWs also affect the spectrum of RGWs in high frequencies.

 Among the current and planed GW detectors,  SKA using the pulsar timing technique and   the   space-based interferometers eLISA, BBO and  DECIGO    are promising to catch the signals of  RGWs. Furthermore, BBO and DECIGO
 have the potential not only to distinguish the spectra with different parameters but also to examine the validity of the quantum normalization. Therefore, RGWs could become the most important tool to know the physics occurred in the very early Universe such as the inflation and reheating process. Even though we chose a series power-law form potential of the scalar field, as shown in \cite{Turner}, a polynomial form of the potential will be dominated by the lowest power of $\phi$ in $V$.  In this case, the conclusion is not substantially modified.
In our previous work \cite{Tong6}, we got the  $r-\beta$ relation for a particular potential  $V= \frac{1}{2}m^2\phi^2$. However, for the more general case $V=\lambda\phi^n$, it is hard to obtain a complete analytic solution of $\trh$ and in turn the increases of the scale factor $\zeta_s$ and $\zeta_1$. Therefore, in this paper, we set a series reasonable values of $\trh$ as additional parameters.
The determination of $n_s$ is very important, since it is sensitively affect our results. The future CMB experiments such as the Plank satellite \cite{Planck}, the ground-based  ACTPol  \cite{Niemack} and the planned CMBPol \cite{CMBpol} will help us to determine the more convincible value of  $n_s$. Therefore, one can expect accordingly that the spectrum of RGWs  would be known better.

In principle, our analysis is valid for the slow-roll inflation with other forms of the potential $V(\phi)$.
However, for some particular forms of $V(\phi)$, it would be difficult to get the analytic
result of $\zeta_1$ as a function of the parameters included in $V(\phi)$. Moreover, one could not effectively constrain the parameters in $V(\phi)$. However, one can still calculate $\zeta_1$ numerically according to the whole calculating processes presented  in Ref.\cite{Mielczarek}, and then calculate the spectra of RGWs accordingly. More general inflationary models other than the slow-roll inflation and
the slow-roll inflation with other forms of $V(\phi)$ would be studied  in our future work.

\

{ACKNOWLEDGMENT}: This work is supported by the National Science Foundation of China  under Grant No. 11103024
and  the  program of the light in China's Western Region of CAS.


\small
\begin{thebibliography}{88}



\bibitem{grishchuk1} L.P. Grishchuk,
     Sov.Phys.JETP {\bf 40}, 409  (1975);
      Class.Quant.Grav.{\bf 14}, 1445 (1997).

\bibitem{grishchuk}   L.P. Grishchuk,
   in {\it Lecture Notes in Physics}, Vol.562, p.167,
   Springer-Verlag, (2001),  arXiv: gr-qc/0002035.
\bibitem{grishchuk3}  L.P. Grishchuk,   arXiv: gr-qc/0707.3319.

\bibitem{starobinsky} A.A. Starobinsky,
        JEPT Lett. {\bf 30}, 682 (1979);
          Sov. Astron. Lett. {\bf 11}, 133  (1985);

V.A. Rubakov, M. Sazhin, and A. Veryaskin,
            Phys. Lett. B {\bf 115}, 189 (1982);

R. Fabbri and M.D. Pollock,
    Phys. Lett. B {\bf125}, 445 (1983);

L. F. Abbott and M.B. Wise,
          Nucl. Phys. B {\bf244}, 541 (1984);



 B. Allen, Phys. Rev. D {\bf37}, 2078 (1988);

V. Sahni,  Phys. Rev. D {\bf42}, 453 (1990);

H. Tashiro, T. Chiba, and  M. Sasaki,
   Class. Quant. Grav. {\bf21}, 1761 (2004);

A. B. Henriques,
    Class. Quant. Grav. {\bf21}, 3057 (2004);

W. Zhao and Y. Zhang,
   Phys. Rev. D {\bf 74},   043503 (2006).



\bibitem{Maggiore}  M. Maggiore, Phys. Rept. {\bf 331}, 283 (2000).
\bibitem{Giovannini} M. Giovannini, PMC Phys. A {\bf4}, 1 (2010).

\bibitem{zhang2} Y. Zhang  {\it et al.},
   Class. Quant. Grav. {\bf 22}, 1383 (2005);
    \bibitem{Zhang4} Y. Zhang  {\it et al.}, Class. Quant. Grav.{\bf 23}, 3783 (2006).

 \bibitem{ligo1} http://www.ligo.caltech.edu/.

\bibitem{ligo2} http://www.ligo.caltech.edu/advLIGO/.

\bibitem{advligo} S. J. Waldman (the LIGO Scientific Collaboration), arXiv:1103.2728.
\bibitem{virgo} A. Freise, {\it et al.},
                      Class. Quant. Grav. {\bf22}, S869 (2005);

                 http://www.virgo.infn.it/.
\bibitem{virgocurve}https://wwwcascina.virgo.infn.it/senscurve/.
\bibitem{geo} B. Willke, {\it et al.}, Class. Quant. Grav. {\bf 19}, 1377 (2002);

              http://geo600.aei.mpg.de/;

              http://www.geo600.uni-hannover.de/geocurves/.

P. Barriga, C. Zhao, and D.G. Blair, Gen. Relativ. Gravit. 37, 1609 (2005).
\bibitem{KAGRA}K. Somiya, Class. Quantum Grav. {\bf29}, 124007 (2012).
\bibitem{Punturo}M. Punturo et al.,  Class. Quantum Grav. {\bf27}, 194002 (2010).
\bibitem{Hild}S. Hild et al.,  Class. Quantum Grav. {\bf28}, 094013 (2011).
\bibitem{lisa0} P. Amaro-Seoane et al, Class. Quantum Grav. {\bf 29, 124016} (2012);
                http://elisa-ngo.org/.
\bibitem{lisa}
         http://www.srl.caltech.edu/\~\,shane/sensitivity/.
\bibitem{Crowder} J. Crowder and N.J. Cornish, Phys. Rev. D, {\bf72}, 083005 (2005).
\bibitem{Cutler}C. Cutler and J. Harms, Phys. Rev. D, {\bf73}, 042001 (2006).
\bibitem{Kawamura}S. Kawamura et al., Class. Quantum Grav. {\bf23}, S125-S131 (2006).
\bibitem{Kudoh}H. Kudoh, A. Taruya, T. Hiramatsu, and Y. Himemoto, Phys. Rev. D, {\bf73}, 064006 (2006).
\bibitem{PPTA}  G. Hobbs,  Class. Quant. Grav. {\bf25}, 114032 (2008);
                        J. Phys. Conf. Ser. {\bf122}, 012003 (2008);

       R.N. Manchester,
                AIP Conf. Series. Proc. {\bf983}, 584 (2008),  arXiv:0710.5026;
                arXiv:1004.3602.
\bibitem{Jenet}
       F.A. Jenet, {\it et al.},  Astrophys. J. {\bf653}, 1571 (2006).
\bibitem{Kramer} M. Kramer et al., New Astr. {\bf48}, 993 (2004);
www.skatelescope.org.
\bibitem{cruise}  A.M. Cruise, Class.Quant.Grav. {\bf 17}, 2525 (2000) ;

         A.M. Cruise and R.M.J. Ingley,
                    Class. Quant. Grav. {\bf 22},  S479 (2005);
                    Class. Quant. Grav. {\bf 23}, 6185 (2006);

         M.L. Tong and Y. Zhang,
                    Chin. J. Astron. Astrophys. {\bf 8}, 314 (2008).



\bibitem{fangyu} F.Y. Li, M.X. Tang and D.P. Shi,
           Phys. Rev. D {\bf 67}, 104008  (2003);

           F.Y.  Li {\it et al}., Eur. Phys. J. C {\bf56}, 407 (2008);

    M.L. Tong, Y. Zhang,  and F.Y. Li,
         Phys. Rev. D {\bf78}, 024041 (2008).

\bibitem{Akutsu} T. Akutsu et al., Phys. Rev. Lett. {\bf101}, 101101 (2008).
\bibitem{basko}  M. Zaldarriaga and  U. Seljak,
                     Phys.Rev.D{\bf 55}, 1830 (1997);

  M. Kamionkowski, A. Kosowsky, and A. Stebbins,
                 Phys. Rev. D{\bf 55}, 7368 (1997);


      B.G. Keating, P.T. Timbie,  A. Polnarev,
                 and  J. Steinberger,  Astrophys. J. {\bf 495}, 580 (1998);

  J. R. Pritchard and  M. Kamionkowski,
                Ann. Phys.(N.Y.) {\bf 318},  2  (2005);

  W. Zhao and  Y. Zhang, Phys.Rev.D{\bf 74}, 083006 (2006);

  T.Y Xia and Y. Zhang, Phys. Rev. D{\bf78},   123005 (2008);
                        Phys. Rev. D{\bf79},  083002 (2009);

   W. Zhao and D. Baskaran, Phys. Rev. D {\bf79}, 083003 (2009).

\bibitem{Peiris} H.V. Peiris, {\it et al},
                       Astrophys. J. Suppl. {\bf148}, 213 (2003).

        D.N. Spergel, {\it et al},
                      Astrophys. J. Suppl. {\bf148}, 175 (2003).


\bibitem{Spergel}  D.N. Spergel, {\it et al},
                        Astrophys. J. Suppl. {\bf170},  377 (2007).

     L. Page, {\it et al},
                        Astrophys.J.Suppl. {\bf 170}, 335 (2007).




\bibitem{Hinshaw} G. Hinshaw,  {\it et al},
                          Astrophys. J. Suppl. {\bf180},  225 (2009);

J. Dunkley,  {\it et al},
                           Astrophys. J. Suppl. {\bf180},  306 (2009).
\bibitem{Komatsu} E. Komatsu, {\it et al},
                          Astrophys. J. Suppl. {\bf192}, 18 (2011).
\bibitem{Planck} Planck Collaboration, arXiv:astro-ph/0604069;
http://www.rssd.esa.int/index.php?project=Planck.
\bibitem{Niemack} M.D. Niemack et al., Proc. SPIE, {\bf7741}, 77411S (2010).

\bibitem{CMBpol}J. Dunkley  {\it et al}., in \emph{CMBPol Mission Concept Study: Prospects for Polarized
Foreground Removal}, 1141, 222 (AIP, New York, 2009).
\bibitem{Kolb} E.W. Kolb and M.S. Turner, \emph{The Early Universe}, (Addison-Wesley, Reading, MA, 1990).
\bibitem{Nakayama} K. Nakayama, S. Saito, Y. Suwa, and J. Yokoyama, JCAP, {\bf0806}, 020 (2008).
\bibitem{Mielczarek} J. Mielczarek, Phys. Rev. D {\bf83}, 023502 (2011).
\bibitem{Tong6} M. Tong, Class. Quantum Grav. {\bf 29}, 155006 (2012).\bibitem{Turner} M.S. Turner, Phys. Rev. D 28, 1243 (1983).
\bibitem{martin} J. Martin and C. Ringeval, Phys. Rev. D {\bf82}, 023511 (2010).
\bibitem{Bailly} S. Bailly, K.-Y. Choi, K. Jedamzik, and L. Roszkowski, J. High Energy Phys. 05, 103 (2009).

\bibitem{Miao} H. X. Miao and Y. Zhang,
                Phys. Rev. D {\bf 75}, 104009 (2007).

\bibitem{Yuki} Y. Watanabe and E. Komatsu, Phys. Rev. D {\bf73}, 123515 (2006).

\bibitem{TongZhang} M.L. Tong, Y. Zhang,
               Phys. Rev. D {\bf 80},  084022  (2009).


\bibitem{Kuroyanagi}S. Kuroyanagi, T. Chiba, and N. Sugiyama, Phys. Rev. D {\bf79}, 103501 (2009).
\bibitem{Starobinsky2} A.A. Starobinsky, Phys. Lett. B {\bf91}, 99 (1980);



\bibitem{zhangtong} Y. Zhang, M.L. Tong, and  Z.W. Fu, Phys. Rev. D   {\bf81}, 101501(R), (2010).
\bibitem{Boyle} L.A. Boyle and P.J. Steinhardt, Phys. Rev. D {\bf77}, 063504 (2008).


\bibitem{gravitinos} M.Y. Khlopov and A.D. Linde, Phys. Lett. B 138, 265 (1984);
C.F. Giudice, I. Tkachev and A. Riotto, J. High Energy Phys. 08, 009 (1999);
M. Lemoine, Phys. Rev. D 60, 103522 (1999);
A.L. Maroto and A. Mazumdar, Phys. Rev. Lett. 84, 1655 (2000);
R. Kallosh, L. Kofman, A.D. Linde, and A. Van Proeyen, Phys. Rev. D 61, 103503 (2000);
A. Buonanno, M. Lemoine, and K.A. Olive, Phys. Rev. D 62, 083513 (2000);
E.J. Copeland and O. Seto, Phys. Rev. D 72, 023506 (2005);
K. Jedamzik, Phys. Rev. D 74, 103509 (2006);
M. Kawasaki, K. Kohri, T. Moroi, and A. Yotsuyanagi, Phys. Rev. D 78, 065011 (2008).


\bibitem{LiddleA}  A. R. Liddle and D. H. Lyth,
             Phys. Lett. B{\bf 291}, 391 (1992);
             Phys. Rep. {\bf231}, 1 (1993);
             {\it Cosmological inflation and large-scale structure},
                   Cambridge University Press (2000).
\bibitem{Kosowsky} A. Kosowsky and M.S. Turner,
                           Phys. Rev. D {\bf 52}, R1739 (1995).



\bibitem{grishchuk91} L.P. Grishchuk and M. Solokhin, Phys. Rev. D {\bf43}, 2566, (1991).

\bibitem{Sanchez}A.G. Sanchez et al, arXiv:1203.6616.
\bibitem{Amarie}M. Amarie, C. Hirata and U. Seljak, Phys. Rev. D {\bf72}, 123006 (2005);

A. Amblard, A. Cooray and M. Kaplinghat, Phys. Rev. D {\bf75}, 083508 (2007).



\bibitem{Boyle2} L.A. Boyle, P.J. Steinhardt, and N. Turok, Phys. Rev. Lett. {\bf96}, 111301 (2006).
\bibitem{Borges} H.A. Borges and S. Carneiro, Gen. Relativ. Gravit. {\bf 37}, 1385 (2005);
                 S. Carneiro, J. Phys. A {\bf 40}, 6841 (2007);
                 M. Tong and  H. Noh, Eur. Phys. J. C {\bf71}, 1586 (2011).
\bibitem{Sesana} A. Sesana, A. Vecchio, and C.N. Colacino, Mon. Not. R. Astron. Soc. {\bf390}, 192 (2008).

\bibitem{Ljz} J. Liu, Mon. Not. R. Astron. Soc. {\bf400}, 1850 (2009); J. Liu, Z. Han, F. Zhang, and Y. Zhang, Astrophys. J. {\bf719}, 1546 (2010); and references therein.



















\end{thebibliography}
\end{document}